# Fuzzy-Based Intelligent Sensors: Modeling, Design, Application

Eric Benoit
LAMII-CESALP,
Université de Savoie
BP 806, F-74016 Annecy cedex
France

Richard Dapoigny
LAMII-CESALP,
Université de Savoie
BP 806, F-74016 Annecy cedex
France

Laurent Foulloy
LAMII-CESALP,
Université de Savoie
BP 806, F-74016 Annecy cedex
France

*Abstract* - **This paper presents a modeling of intelligent sensors based on a representation of the sensor by services it uses or it proposes, and by its USer Operating Modes (USOMs). This modeling is used for the definition of the reactive layer of distributed agent based intelligent sensors. Our area of interest is the agent-level layer in which the concept of IIC (Intelligent Instrument Cluster) is defined. An application that uses fuzzy-based intelligent sensors is presented in order to illustrate the concepts.**

## I. INTRODUCTION

Since the eighties, sensors have evolved toward more complex functionalities. They now include functionalities such as communication, configuration and validation in addition to measurement ones. The introduction of the intelligence concept into sensors has enabled to consider higher functionnalities like learning, self-diagnosis or group decision. When applied to sensors, intelligence can be declined into three large categories: intelligence of the perception, reasoning, and social intelligence. The intelligence of the perception denotes the ability to understand the physical environment. Reasoning denotes the ability to create new knowledge from the existing one. The social intelligence denotes the ability to exchange knowledge with other actors.

In order to implement intelligence into sensors, two kinds of modeling are needed. The first one includes information modeling. It covers research fields like the probability theory, the possibility theory, the measurement theory, fuzzy sensors, symbolic sensors, uncertainty management. In order to increase sensor functionalities, symbolic sensors have been introduced [1]. Unlike a numeric sensor that provides an objective quantitative description of the objects, a symbolic sensor provides a subjective qualitative description of objects. This qualitative description, adapted to the sensor measurement, can be used in knowledge based decision systems, for example for checking the validity of a measurement, or improving the relevance of a result. Fuzzy techniques have been developed in order to model measurement information, both for numeric measurements and symbolic measurements. Two families of fuzzy sensors are differentiated [2]. Fuzzy symbolic sensors provide fuzzy subsets of linguistic terms [3]. They are used preferentially to implement human-like perception [4], [5]. Fuzzy numeric sensors provide fuzzy subsets of numeric values that can represent a possibility distribution.

The second kind of modeling includes behaviour modeling like object modeling, functional modeling, or event-based modeling. A well defined modeling allows to study original concepts like subcontractor computing units [6], or processor free intelligent sensors. This paper presents a distributed agent modeling of sensors. It is an extension to a service-based approach that splits the sensor behavior into internal modes (INOM) [7]. In this modeling, services are included in three layers. The lower layer presents basic services that deal with sensor hardware and basic algorithms. The second level has services that can be requested by an outer entity. The higher layer is defined as a distributed agent layer.

## II. SERVICE-BASED INTELLIGENT SENSOR MODELING

Recent studies propose models based on a set of functionalities organized with a general behavioral description, i.e. automation graph or object mode [8][9][10][11]. The internal modeling of intelligent instruments is not sufficient for the design of large applications. Obviously, intelligent instruments need to inter-operate. Therefore an external model of intelligent instrument is required. In [12] and [13], Staroswiecki proposed to model a sensor by a set of services. Services are organized into subsets called "USer Operating Modes". In this model, a sensor service can be requested, and so serviced, only if the current active User Operating Mode (USOM) includes this service. This prevents the requirement of services when they cannot be available.

The approach discussed in [12] was proposed to model existing instruments from the external point of view. In particular, the external model of the instrument can be used to build a global model for an application involving several instruments. This kind of approach can also be used to define the internal functional model of a sensor.

Instruments are considered as entities that offer some more or less complex services. These services represent the instrument functionalities from the user's point of view. At a lower level, each instrument service is defined as a set of internal services.

These two levels are representative of the gap between the instrument user point of view, the instrument designer point of view and the software designer activity. In order to use instrument designer capabilities for the design of intelligent instruments, instrument functionalities are described with basic internal services. Then the designer will just have to define each external service with a set of internal services.

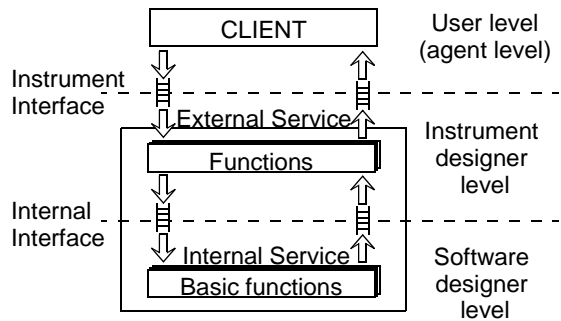

Fig. 1. Service based structure of an intelligent instrument

The last level, called the "user level", is initialy devoted to the management of service request. In the approach presented in this paper, this level is devoted to the agent activity. The interface between the agent and the instrument is included into the external model of the instrument. It is composed of:

- a set of external services. Any external service is a service that can be required by any entity of the user level.

- a set of external events. An external event can be a service request event, it is performed at the user level.

- a mapping that links external services and external events. If an external service $es$ and an external event $e$ are related, then the occurence of the event $e$ activates the service $es$.

- a set of USOMs. each USOMs is a subset of external services.

- a relation on USOMs that represents admitted transitions between USOMs. This relation can be seen as a simple oriented graph.

- a function that links mode transitions and external events. If a transition $(m1,m2)$ and an external event $e$ are related, then the occurence of the event $e$ switches the current mode from $m1$ to $m2$.

From the agent point of view, the local instrument can receive events that activate services or switch modes. It can give its current mode. This modeling allows the agent to know if a service can be activated or not, and if a mode transition is allowed or not.

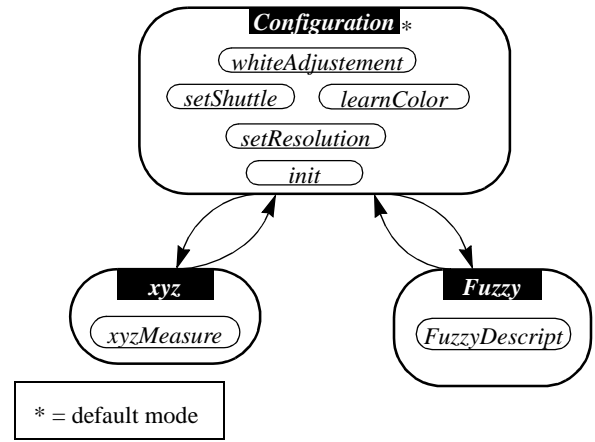

Fig. 2. Exemple of external modes and external services of an intelligent spectrometer

This modeling can be used to simplify the sensor design by an automatic generation of the sensor software and the testing software [14]. A set of tools associated to this modeling performs model checking, consistency checking and software generation. It guarantees the consistency of the sensor behaviour.

### III. DISTRIBUTED AGENT APPROACH

*A. Intelligent Agent Architecture*

Intelligent Instrument modelling (USOM) defines an entity offering external services which handle variables and call for a set of resources. In a distributed model, the User Mode concept is not sufficient if instruments must cooperate in a global application. Recent works have expressed the distributed application in terms of a graph topology which represents the asynchronous product of the individual graphs of each instrument under constraints. Constraints are materialized by events conditionning external service activity [15]. The following method is based upon intelligent agent concept, which leads to a dynamic behaviour rather than a static behaviour.

In distributed artificial intelligence (AI), agents are small autonomous units which are able to perceive, to exchange information with others, to plan, to decide and to act. Perception and action are associated with the input/output process. Planning and deciding are implicitly programmed in the kernel of the intelligent instrument [16]. More precisely, agents in a distributed system are entities which have some key properties [17] :

- *autonomy* : agents have control over their actions and internal state.

- *sociability* : agents are able to cooperate with other agents (or humans) to achieve their tasks.

- *reactivity* : agents perceive their environment and respond to changes that occur in it.

- *proactiveness* : agents are able to exhibit goal-directed behavior by taking the initiative.

Globally, agent-based computing is divided into two levels, the agent-level layer which maps the properties of autonomy, reactivity, proactiveness and social ability at the lower level and the societal level which focuses on the distributed AI aspects, i.e. cooperation, coordination and negotiation.

Our area of interest is the agent-level layer in which the concept of IIC (*Intelligent Instrument Cluster*) is defined. An IIC is defined as an autonomous and cooperative subsystem made of several nodes, where each node is seen as an intelligent agent. In practise, the IIC is a network of embedded systems where each node is arranged with a microcontroller connected to a smart sensor/actuator with large pre-processing/self-control facilities. The system automatically realizes the communications between the agents through the field-bus.

*B. The Agent-based Model*

The following agent-based model described below is derived from the hybrid architecture which associates the two classical types, i.e. the symbolic and the reactive architectures [18]. Earlier works in the area of agent architectures have established that the most popular approach to design hybrid architectures are the BDI approach [19] and the layered architecture [20]. The BDI architecture typically contains four data structures, i.e. beliefs, desire intentions and plans which are linked to an interpreter. The layered architecture allows us to integrate different agent control subsystems by layering them. Our model derives from the BDI architecture for the following reasons :

- it is the best-known agent architecture,

- it maps roughly the service-based approach to intelligent instrument design,

- it gives a generic approach which can be easily improved by adding new features.

In the service-based intelligent sensor modeling, a Man-Machine Interface has been developed [14]. The user defines a set of possible services and starts the instrument in an initial state The intelligence of the instrument lies in the ability of evolving through different modes according to the validity of results. The intelligent instrument is able to adapt and to deal with sensor measurements without human interference.

Our BDI-based model relies on a functional reasoning node. The agent beliefs correspond to information the agent has about its environment. This information can be represented by exported variables or external services broadcasted from other instruments. There is a general trend in designing distributed intelligent instruments to give an increasing amount of autonomy to the individual nodes of such architectures. Recent studies [21] show that a model where intelligent nodes are coupled by well-defined event channels preserve autonomy and identify typical situations arising in distributed control applications (i.e. the information produced at one place is relevant for a number of consumers). Service requirements are event driven and information gained from a sensor can be used and analysed in more than one node. The agent desires correspond to the tasks allocated to the node, i.e. the local available states which must be consistent with the others. The agent intentions represent external services that the agent is committed to achieve in the current mode.

The plan usually contains two parts, a program and a descriptor. The first defines a course of actions. The latter states both the circumstances under which the plan can be used and what intentions the plan may use in order to achieve its goal. This function is performed by the internal and external service description downloaded into each node. The kernel acts as an interpreter which selects an action to perform (i.e. services to perform) on the basis of the agent current intentions and procedural knowledge (i.e. the available services at a given time). Figure 3 illustrates the major functional elements and interfaces of the functional reasoning node.

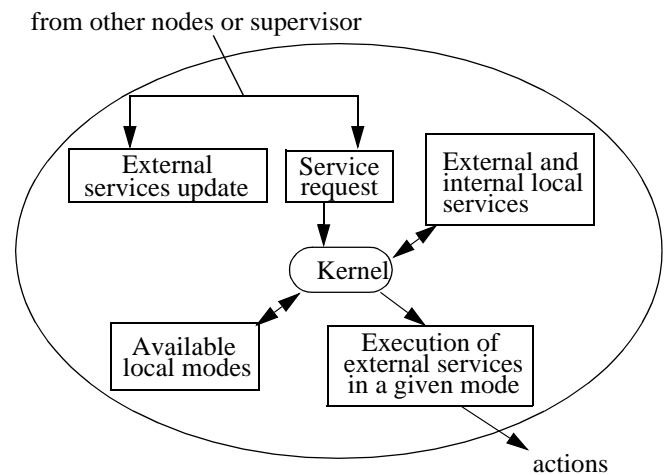

Fig. 3. The functional reasoning model.

*C. Multi-Agent Implementation*

In this environment, we operate at the intermediary level between the cognitive and reactive level. The agent tends to fulfill its goals with taking its local resources and its internal operating mode into account and considering its communications. Communications are performed via message exchange.

A variable is the basic unit transmitted between external services on each instrument. Thus, the environment knowledge is implemented by updating a local database into each instrument. This database, called the *external representation*, is divided into several fields such as node number, variable name, variable local and external numbers... In the previous model, no relation were existing between variables on differents nodes. The main hypotheses are that:

- a variable is generated by one and only one node,

- a variable name is common to all parts of a distributed application (i.e. a same name represents a same entity).

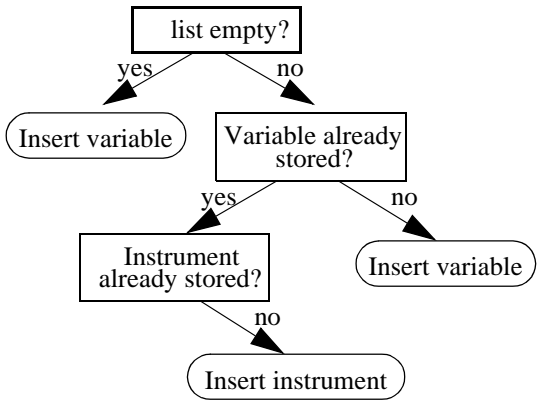

Fig. 4. The update protocol algorithm

When a node starts in the network, it broadcasts its local exported variables list to every node of the IIC. Each node responds by sending its exported variable list. This exchange is performed at starting time for the new instrument, and during execution time for the other instruments. After this broadcast step, each database is updated making the variable representation independant from any instrument.

First, this method enhances redundancy since many instruments can cooperate, each of them producing a variable as a result from the same measurement. Reliability can be extended as more as the user wants.

Secondly, each instrument contributes to execute a portion of the global application goal. As each instrument implements the same model, its specialization (due to the sensor/actuator component connected to it) reflects the modularity of this kind of application.

Third, as instruments can interact, they can evolve through different modes and activate different external services according to the actual operating mode. This illustrates the adaptativity resulting from the above method. For example, a local node including a spectrometer computes the color of the actual sample in front of a sample holder (see application below). If the result doesn't belong to a pre-determined interval, the instrument is switched to a degraded mode in which an external service can send a given variable to another instrument connected to an actuator (step-in motor) which can move the sample according to the value which has been transmitted.

Model-based instrument architecture using the functional reasoning model described above will provide superior modularity and transparence allowing for easier sensor fusion and knowledge extraction.

## IV. APPLICATION EXAMPLE

*A. Goal*

The application example chosen to illustrate this paper uses a spectrometer [22], a mechanized sample holder, and a telemeter. These three intelligent instruments are communicating through a CAN network [23]. The aim of this application is on-line spectrum analyses of samples. An independent webcam allows the final user to monitor the application. The telemeter is used to detect if the application is isolated i.e. if no operator are manipulating the sample holder.

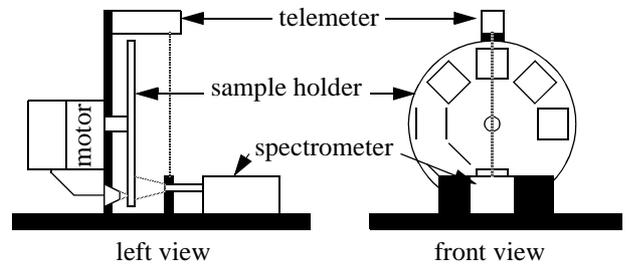

Fig. 5. Mechanical view of the application example

*B. Node Description*

The holder can hold seven samples. The place for a eighth sample (sample 0) is kept empty in order to allow the learning of the color of the holder background. The holder general behaviour, shown in figure 6, distinguish a *manual* mode that includes direct angle modification services and a *sample* mode that includes sample selection services.

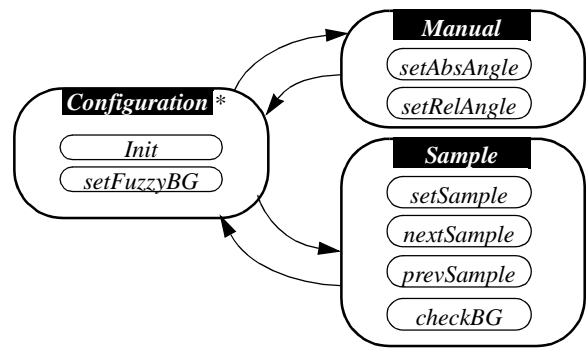

Fig. 6. External model of the sample holder

- The *init* service makes the holder searching for the initial position. It also sends learning data for the telemeter: *distanceExample* and *fuzzyDistanceExample* (see table I).

- The *setFuzzyBG* service stores the variable *fuzzyBGValue* that represents the fuzzy description of the background measurement (see table II).

- The services *setAbsAngle* and *setRelAngle* makes the sample holder turning its disk. The real angle is sent as acknowledgment of the movement.

- The *setSample* service put the sample in front of the color sensor. The number is into the *sampleNumber* local variable.

- Services *nextSample* and *prevSample* switch between samples. The sample number is sent as acknowledgment of the movement.

- The service *checkBG* gets the fuzzy description of the current sample measurement (*fuzzySampleValue*) and compares it with the stored background description. If the background is recognized, the sample holder repeats the last movement.

The spectrometer can perform a classical measurement of xyz chromaticity coordinates or a fuzzy description of the color. The spectrometer initially performs a fuzzy description on the universe of discourse: {*red*, *green*, *blue*, *yellow*, *cyan*, *magenta*, *grey*}.

- In the *Configuration* mode, the *learnColor* service uses the measured xyz chromaticity coordinates and the *colorNumber* in order to improve the perceptive knowledge of the color sensor.

- During the *FuzzyDescript* external service, the sensor sends the *xyz* and the fuzzy description of the *fuzzyColor* variable.

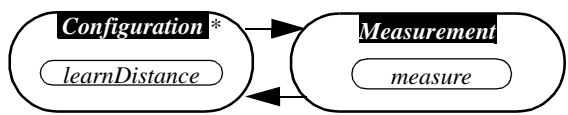

Fig. 7. External model of the telemeter

The telemeter is able to give a numerical representation of the distance and a fuzzy linguistic representation of the same distance.

- The *measure* telemeter service sends the numerical value of the distance, and the fuzzy description of the distance.

- The *learnDistance* service uses the numerical value (*exampleValue*) of a distance example and it is associated fuzzy description (*fuzzyExample*) in order to build the perceptive knowledge of the telemeter. If the local variable *exampleValue* is not linked with an external variable, the telemeter acquires itself the distance.

TABLE I
LIST OF EXPORTED VARIABLES

| source node | Name | type |
| --- | --- | --- |
| Sample holder | angle | unsigned integer 16 bits |
| Sample holder | sampleNumber | byte |
| Sample holder | distanceExample | unsigned integer 16 bit |
| Sample holder | fuzzyDistExample | fuzzy subset |
| color sensor | xyz | table of 3 bytes |
| color sensor | fuzzyColor | fuzzy subset |
| telemeter | distance | unsigned integer 16 bits |
| telemeter | fuzzyDistance | fuzzy subset |

TABLE II
LIST OF IMPORTED VARIABLES

| consumer node | Local name | type |
| --- | --- | --- |
| Sample holder | angle | unsigned integer 16 bits |
| Sample holder | sampleNumber | byte |
| Sample holder | fuzzySampleValue | fuzzy subset |
| Sample holder | fuzzyBGValue | fuzzy subset |
| telemeter | fuzzyExample | fuzzy subset |
| telemeter | exampleValue | unsigned integer 16 bits |
| color sensor | colorNumber | byte |

*C. Definition of the Application*

The definition of the application first starts by establishing the data-flow at the application level, i.e. defining the links between exported variables and local imported variables (fig. 8). The non linked variables can be directly defined by an external user.

The behaviour can be defined in the same way, i.e. with the links between exported events like variable emission or mode switching, and imported events like service request or mode switching request.

In this example, a reactive behaviour is defined. The color sensor gives a fuzzy description each time a sample is in front of it. And the sample holder skips a sample place each time the color sensor gives the fuzzy description of the background. It is also possible to define a reactive behavior that switches the sample holder and the color sensor into their *Configuration* modes when the telemeter detects an external object. In order to define this behaviour, a new service *checkDistance* has to be defined. An other way is to consider that the decision has to be taken in a higher level. The localisation of the decision process must be dicussed before improving the approach presented into this paper.

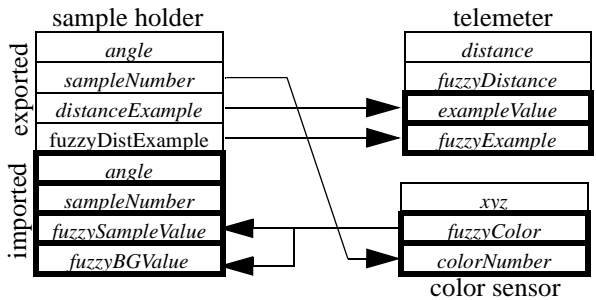

Fig. 8. Links between exported variables and local imported variables

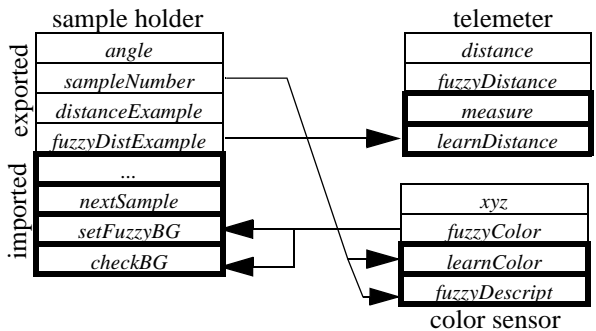

Fig. 9. Links between events (switching-mode events are not presented)

## V. CONCLUSION

Our BDI-based model will extend the service-based intelligent sensor modeling to a more global approach where a set of intelligent instruments can cooperate to enhance their knowledge, and to understand each other in order to achieve common goals. From this point of view, human operators will be integrated as ultimate strategic-level decision makers. Future works will extend the method to external services knowledge to construct a dependency network.